\def\year{2015}
\documentclass[letterpaper]{article}
\usepackage{aaai}
\usepackage{times}
\usepackage{helvet}
\usepackage{courier}
\usepackage[hyphens]{url}
\usepackage{graphicx,subfigure}
\begin{document}
%
\title{Detecting Malicious Content on Facebook}
\author{Prateek Dewan, Ponnurangam Kumaraguru\\
Indraprastha Institute of Information Technology - Delhi (IIITD), India\\
Cybersecurity Education and Research Centre (CERC)\\
\{prateekd, pk\}@iiitd.ac.in\\}

\maketitle
\begin{abstract}
\begin{quote}

Online Social Networks (OSNs) witness a rise in user activity whenever an event takes place. Malicious entities exploit this spur in user-engagement levels to spread malicious content that compromises system reputation and degrades user experience. It also generates revenue from advertisements, clicks, etc. for the malicious entities. Facebook, the world's biggest social network, is no exception and has recently been reported to face much abuse through scams and other type of malicious content, especially during news making events. Recent studies have reported that spammers earn \$200 million just by posting malicious links on Facebook. In this paper, we characterize malicious content posted on Facebook during 17 events, and discover that existing efforts to counter malicious content by Facebook are not able to stop all malicious content from entering the social graph. 
Our findings revealed that malicious entities tend to post content through web and third party applications while legitimate entities prefer mobile platforms to post content. In addition, we discovered a substantial amount of malicious content generated by Facebook pages. 
Through our observations, we propose an extensive feature set based on entity profile, textual content, metadata, and URL features to identify malicious content on Facebook in real time and at zero-hour. This feature set was used to train multiple machine learning models and achieved an accuracy of 86.9\%. The intent is to catch malicious content that is currently evading Facebook's detection techniques. 
Our machine learning model was able to detect more than double the number of malicious posts as compared to existing malicious content detection techniques. Finally, we built a real world solution in the form of a REST based API and a browser plug-in to identify malicious Facebook posts in real time.

\end{quote}
\end{abstract}

\section{Introduction}

Social network activity rises considerably during events that make the news, like sports, natural calamities, etc.~\cite{szell2014}. The FIFA World Cup in 2014, for example, saw a record-breaking 350 million users generating over 3 billion posts on Facebook over a period of 32 days.~\footnote{\url{http://edition.cnn.com/2014/07/14/tech/social-media/world-cup-social-media/}} Such colossal magnitude of activity makes OSNs an attractive venue for malicious entities. Facebook, world's biggest social network, is no exception. Being the most preferred OSN for users to get news~\cite{Jesse-Holcomb:2013}, Facebook is potentially one of the most lucrative OSNs for malicious entities. Recently, a group of malicious users exploited the famous biting incident during the 2014 FIFA World Cup, where an Uruguayan player was banned for biting an opponent. Attackers used the viral nature of this incident to spread links on Facebook, pointing to a phishing page prompting visitors to sign a petition in defense of the Uruguayan player.~\footnote{\url{http://www.emirates247.com/news/fifa-world-cup-2014-smitten-by-suarez-bitten-by-spammers-2014-07-12-1.556216}} The petition required a user to sign in with details such as name, country of residence, mobile phone number and email address.
The petitioner could potentially end up on a spam mailing list, on the receiving end of a malicious attachment or even subjected to a targeted attack. In another recent incident of Malaysian Airline MH17 flight crash, scammers placed dozens of so-called `community pages' on Facebook, dedicated to victims of the tragedy. On the page, Facebook users were tricked into clicking links showing extra or unseen footage of the crash. Instead of seeing a video, they were led to various pop-up ads for porn sites or online casinos.~\footnote{\url{http://www.nltimes.nl/2014/07/22/flight-17-spam-scams-facebook-twitter/}} Such activity not only violates Facebook's terms of service, but also degrades user experience. It has been claimed that Facebook spammers make \$200 million just by posting links.~\footnote{\url{http://www.theguardian.com/technology/2013/aug/28/facebook-spam-202-million-italian-research}} 
Facebook has confirmed spam as a serious issue, and taken steps to reduce spam and malicious content in users' newsfeed recently~\cite{Owens:2014}. 

The problem of identifying malicious content is not specific to Facebook and has been widely studied on other OSNs in the past. Researchers have used feature based machine learning models to detect spam and other types of malicious content on OSNs like Twitter, and achieved good results~\cite{benevenuto2010detecting,grier2010spam}. However, existing approaches to detect malicious content in other OSNs like Twitter, cannot be directly ported to Facebook because they heavily rely on features that aren't publicly available from Facebook. These include profile, and network information, age of the account, total number of messages posted, number of social connections, etc. 

In this paper, we highlight that existing techniques used by Facebook for countering malicious content do not eliminate all malicious posts completely. Although Facebook's immune system~\cite{stein2011facebook} seems to perform well at protecting its users from malicious content, our focus is on detecting the fraction of content which evades this system. We identify some key characteristics of malicious content spread on Facebook, which distinguishes it from legitimate content. Our dataset consists of 4.4 million public posts generated by 3.3 million unique entities during 17 events, across a 16 month time frame (April 2013 - July 2014). We then propose an extensive set of 42 features which can be used to distinguish malicious content from legitimate content in real time and at zero-hour. We emphasize on zero-hour detection because content on OSNs spreads like wildfire, and can reach thousands of users within seconds. Such velocity and reach of OSN content makes it hard to control the spread of malicious content, if not detected instantly. We apply machine learning techniques to identify malicious posts on Facebook using this feature set and achieve a maximum accuracy of 86.9\% using the Random Forest classifier. We also compare our technique with past research and find that our machine learning model is able to detect more than twice the number of malicious posts as compared to clustering based campaign detection techniques used in the past. 


\section{Related work}

Facebook has its own immune system~\cite{stein2011facebook} to safeguard its users from unwanted malicious content. Researchers at Facebook built and deployed a coherent, scalable, and extensible real time system to protect their users and the social graph. This system performs real time checks and classifications on every read and write action. 
Designers of this complex system used an exhaustive set of components and techniques to differentiate between legitimate actions and spam. These components were standard classifiers like Random Forest, Support Vector Machines, Logistic Regression, a feature extraction language, dynamic model loading, a policy engine, and feature loops. Interestingly, despite this complex immune system deployed by Facebook, unwanted spam, phishing, and other malicious content continues to exist and thrive on Facebook. Although the immune system deployed by Facebook utilizes a variety of techniques to safeguard its users, authors did not present an evaluation of the system 
to suggest how accurately and efficiently the system is able to capture malicious content. 

\paragraph{Detection of malicious content on Facebook}

Gao et al., in 2010, presented an initial study to quantify and characterize spam campaigns launched using accounts on Facebook~\cite{gao2010detecting}. They studied a large anonymized dataset of 187 million asynchronous ``wall" messages between Facebook users, and used a set of automated techniques to detect and characterize coordinated spam campaigns. Authors detected roughly 200,000 malicious wall posts with embedded URLs, originating from more than 57,000 user accounts. 
Following up their work, Gao et al. presented an online spam filtering system that could be deployed as a component of the OSN platform to inspect messages generated by users in real-time~\cite{gao2012towards}. Their approach focused on reconstructing spam messages into campaigns for classification rather than examining each post individually. They were able to achieve a true positive rate of slightly over 80\% using this technique, and achieved an average throughput of 1,580 messages/sec with an average processing latency of 21.5ms on their Facebook dataset of 187 million wall posts. However, the clustering approach used by authors always marked a new cluster as non malicious, and was unable to detect malicious posts if the system had not seen a similar post before. We overcome this drawback in our work by eliminating dependence on post similarity, and using classification instead of clustering. 

In an attempt to protect Facebook users from malicious posts, Rahman et al. designed an efficient social malware detection method which took advantage of the social context of posts~\cite{rahman2012efficient}. Authors were able to achieve a maximum true positive accuracy rate of 97\%, using a SVM based classifier trained on 6 features, and requiring 46 milliseconds to classify a post. This model was then used to develop MyPageKeeper~\footnote{\url{https://apps.facebook.com/mypagekeeper/}}, a Facebook app to protect users from malware and malicious posts. Similar to Gao et al's work~\cite{gao2010detecting}, this work was also targeted at detecting spam campaigns, and relied on message similarity features. 

\paragraph{Detection of malicious content on other OSNs}

Multiple machine learning based techniques have been proposed in the past to detect malicious content on other social networks such as Twitter and YouTube~\cite{benevenuto2010detecting,grier2010spam,wang2010don,mccord2011spam}. The efficiency of such techniques comes from features like age of the account, number of social connections, past messages of the user, etc.~\cite{benevenuto2010detecting}. However, none of these features are available on Facebook publicly. Other techniques make use of OSN specific features like user replies, user mentions, retweets (Twitter)~\cite{grier2010spam}, post views and ratings (YouTube)~\cite{benevenuto2009detecting} to identify malicious content, which cannot be ported to Facebook. Blacklists have been shown to be highly ineffective initially, capturing less than 20\% URLs at zero-hour~\cite{sheng2009empirical}. None of these techniques can thus be used for efficient zero-hour detection of malicious content on Facebook. To the best of our knowledge, our technique is one of the first attempts towards zero-hour detection of malicious content on Facebook. 
\\


Most aforementioned techniques to identify malicious content on Facebook largely rely on message similarity features. Such techniques are reasonably efficient in detecting content which they have seen in the past, for example, campaigns. However, none of the these techniques are capable of detecting malicious posts which their systems haven't seen in the past. Researchers have acknowledged this drawback~\cite{gao2010detecting} and zero-hour detection of malicious content on Facebook remains an unaddressed problem.


\section{Methodology}



There exists a wide range of malicious content on OSNs today. These include phishing URLs, spreading malware, advertising campaigns, content originating from compromised profiles, artificial reputation gained through fake likes, etc. We do not intend to address all such attacks. We focus our analysis on identifying posts containing one or more malicious URLs and creating automated means to detect such posts in real time, without looking at the landing pages of the URLs. We emphasize on not visiting the landing pages of URLs since this process induces time lag and increases the time taken by real time systems to make a judgment on a post. 
Existing methods involve detection of such malicious posts by grouping them into campaigns~\cite{gao2010detecting}, or by looking up public blacklists like PhishTank, Google Safebrowsing, etc. to identify malicious URLs. However, as previously discussed, both campaign detection techniques and URL blacklists prove ineffective while the attack is new. 
For the scope of this work, we refer to a post as malicious if it contains one or more malicious URLs. 

\subsection{Data collection} \label{sec:data}

We collected data using Facebook's Graph API Search endpoint~\cite{Facebook-Developers:2013} during 17 events that took place between April 2013 and July 2014. The search endpoint in version 1.0 of the Graph API allows searching over many public objects in the social graph, like users, posts, events, pages etc., using a search query keyword. We used event specific terms for each of the 17 events (see Table~\ref{tab:events_keywords}) to collect relevant public \emph{posts}. Unlike other social networks like Twitter, Facebook does not provide an API endpoint to collect a continuous random sample of public posts in real time. Thus, we used the search API to collect data. A drawback of the search method is that if a post is deleted or removed (either by the user herself, or by Facebook) before our data collection module queries the API, it would not appear in the search results. We repeated the search every 15 minutes to overcome this drawback to some extent. In all, we collected over 4.4 million public Facebook posts generated by over 3.3 million unique entities. Table~\ref{tab:descstats} shows the descriptive statistics of our final dataset.

\begin{table*}[!ht]
\small
\begin{centering}
    \begin{tabular}{p{6.6cm}|p{1.1cm}|p{8.8cm}}
    \hline
 Event \texttt{(keywords)}         & \# Posts             & Description  \\ \hline
\raggedright Missing Air Algerie Flight AH5017 \texttt{(ah5017; air algerie)} & 6,767   &  Air Algerie flight 5017 disappeared from radar 50 minutes after take off on July 24, 2014. Found crashed near Mali; no survivors.                  \\ \hline
\raggedright Boston Marathon Blasts \texttt{(prayforboston; marathon blasts; boston marathon)} & 1,480,467 & Two pressure cooker bombs exploded during the Boston Marathon at 2:49 pm EDT, April 15, 2013, killing 3 and injuring 264.\\ \hline
\raggedright Cyclone Phailin \texttt{(phailin; cyclonephailin)}   & 60,016 & Phailin was the second-strongest tropical cyclone ever to make landfall in India on October 11, 2013.    \\ \hline
\raggedright FIFA World Cup 2014 \texttt{(worldcup; fifaworldcup)}    & 67,406  & 20th edition of FIFA world cup, began on June 12, 2014. Germany beat Argentina in the final to win the tournament. \\ \hline
\raggedright Unrest in Gaza \texttt{(gaza)}  & 31,302   & Israel launched Operation Protective Edge in the Hamas-ruled Gaza Strip on July 8, 2014.     \\ \hline
\raggedright Heartbleed bug in OpenSSL \texttt{(heartbleed)}  & 8,362  & Security bug in OpenSSL disclosed on April 1, 2014. About 17\% of the world's web servers found to be at risk.      \\ \hline
\raggedright IPL 2013 \texttt{(ipl; ipl6; ipl2013)}  &   708,483  &  Edition 6 of IPL cricket tournament hosted in India, April-May 2013.    \\ \hline
\raggedright IPL 2014 \texttt{(ipl; ipl7)} &   59,126  & Edition 7 of IPL cricket tournament jointly hosted by United Arab Emirates and India, April-May 2013.   \\ \hline
\raggedright Lee Rigby's murder in Woolwich \texttt{(woolwich; londonattack)} & 86,083  & British soldier Lee Rigby attacked and murdered by Michael Adebolajo and Michael Adebowale in Woolwich, London on May 22, 2013.          \\ \hline
\raggedright Malaysian Airlines Flight MH17 shot down \texttt{(mh17)}  & 27,624    &  Malaysia Airlines Flight 17 crashed on 17 July 2014, presumed to have been shot down, killing all 298 on board.     \\ \hline
\raggedright Metro-North Train Derailment \texttt{(bronx derailment; metro north derailment; metronorth)} & 1,165 & A Metro-North Railroad Hudson Line passenger train derailed near the Spuyten Duyvil station in the New York City borough of the Bronx on December 1, 2013. Four killed, 59 injured. \\ \hline
\raggedright Washington Navy Yard Shootings \texttt{(washington navy yard; navy yard shooting; NavyYardShooting)} & 4,562   & Lone gunman Aaron Alexis killed 12 and injured 3 in a mass shooting at the Naval Sea Systems Command (NAVSEA) headquarters inside the Washington Navy Yard in Washington, D.C. on Sept. 16, 2013. \\ \hline
\raggedright Death of Nelson Mandela \texttt{(nelson; mandela; nelsonmandela; madiba)} & 1,319,783 & Nelson Mandela, the first elected President of South Africa, died on December 5, 2013. He was 95. \\ \hline
\raggedright Birth of the fist Royal Baby \texttt{(RoyalBabyWatch; kate middleton; royalbaby)} & 90,096  & Prince George of Cambridge, first son of Prince William, and Catherine (Kate Middleton), was born on July 22, 2013.  \\ \hline
\raggedright Typhoon Haiyan \texttt{(haiyan; yolanda; typhoon philippines)}  & 486,325  &  Typhoon Haiyan (Yolanda), one of the strongest tropical cyclones ever recorded, devastated parts of Southeast Asia on Nov. 8, 2013.     \\ \hline
\raggedright T20 Cricket World Cup \texttt{(wt20; wt2014)}      & 25,209  & Fifth ICC World Twenty20 cricket competition, hosted in Bangladesh during March-April, 2014. Sri Lanka won the tournament.         \\ \hline
\raggedright Wimbledon Tennis 2014 \texttt{(wimbledon)}     & 2,633  &    128th Wimbledon Tennis championship held between June 23, and July 6, 2014. Novak Djokovic from Serbia won the championship.    \\ \hline
    \end{tabular}
\vspace{-10pt}
\caption{Event name, keywords used as search queries, number of posts, and description for each of the 17 events in our dataset.}
\label{tab:events_keywords}
\vspace{-10pt}
\end{centering}
\end{table*}

\paragraph{Event selection}

All events we picked for our analysis, made headlines in international news. To maintain diversity, we selected events covering various domains of news events like political, sport, natural hazards, terror strikes and entertainment news. We tried to pick a mixture of crisis, and non crisis events which took place in, and affected different parts of the world. In terms of data, we selected events with at least 1,000 public Facebook posts. For all 17 events, we started data collection from the time the event took place, and stopped about two weeks after the event ended. 

%
%
%

%
%
%
%
%

\subsection{Labeled dataset creation} \label{sec:labeled_dataset}

To create a labeled dataset, we first filtered out all posts containing at least one URL. We used regular expressions to filter out all possible URLs from the \emph{message} field of the posts. These URLs were added to the set of URLs present in the \emph{link} field (if available) for each post. These extracted URLs were then visited using Python's Requests package~\footnote{\url{http://docs.python-requests.org/en/latest/}} and LongURL API~\footnote{\url{http://longurl.org/api}}, in case the Requests package failed. Visiting the landing pages of the URLs before the blacklist lookup helped us to eliminate invalid URLs and capture the correct final destination URLs corresponding to shortened URLs, if any. After the extraction and validation process, we were left with a total of 480,407 unique URLs across 1,222,137 unique posts (Table~\ref{tab:descstats}). Each URL was then subjected to five blacklist lookups, viz. SURBL~\cite{surbl2011reputation}, Google Safebrowsing~\cite{Google:2014}, PhishTank~\cite{OpenDNS:2014}, VirusTotal~\cite{Hispasec-Sistemas-S.L.:2013}, and Web of Trust~\cite{WOT:2014}. This methodology of identifying malicious content using URL blacklists has also been used multiple times in past research~\cite{gao2010detecting,chu2012detecting,thomas2011design}.

\begin{table}[!h]
\begin{center}
\small
    \begin{tabular}{l|r}
    \hline
    Unique posts                        & 4,465,371       \\ \hline
    Unique entities                        & 3,373,953        \\
    - Unique users                           & 2,983,707        \\
    - Unique pages                         &  390,246          \\ \hline
    Unique URLs                          & 480,407          \\  \hline
    Unique posts with URLs              & 1,222,137        \\ \hline
    Unique entities posting URLs           &          856,758 \\ \hline
    Unique posts with malicious URLs    &          11,217  \\ \hline
    Unique entities posting malicious URLs &          7,962   \\ \hline
    Unique malicious URLs               & 4,622            \\ \hline
    \end{tabular}
\caption{Descriptive statistics of complete dataset collected over April 2013 - August 2014.}
\label{tab:descstats}
\end{center}
\vspace{-15pt}
\end{table}

The scan results returned by the VirusTotal API contain domain information from multiple services like TrendMicro, BitDefender, WebSense ThreatSeeker, etc. for a given domain. We marked a URL as malicious if one or more of these services categorized the domain of the URL as \emph{spam}, \emph{malicious}, or \emph{phishing}. The Web of Trust (WOT) API returns a reputation score for a given domain. Reputations are measured for domains in several \emph{components}, for example, trustworthiness. For each  {\fontfamily{qcr}\selectfont \{domain, component\}} pair, the system computes two values: a \emph{reputation} estimate and the \emph{confidence} in the reputation. Together, these indicate the amount of trust in the domain in the given component. A \emph{reputation} estimate of below 60 indicates \emph{unsatisfactory}. Also, the WOT browser add-on requires a confidence value of $\geq$ 10 before it presents a warning about a website. We tested the domain of each URL in our dataset for two components, viz. \emph{Trustworthiness} and \emph{Child Safety}. For our experiment, a URL was marked as malicious if both the aforementioned conditions were satisfied. That is, if the \emph{reputation} estimate for a domain of a URL was below 60 (unsatisfactory, poor or very poor) and the \emph{confidence} in the reputation was $\geq$ 10, the URL was marked malicious. In addition to reputations, the WOT rating system also computes categories for websites based on votes from users and third parties. We marked a URL as malicious if it fell under the \emph{Negative} or \emph{Questionable} category group.~\footnote{The exact category labels corresponding to \emph{Negative} and \emph{Questionable} categories can be found at \url{https://www.mywot.com/wiki/API}} Further, a URL was marked malicious if it was present under any category (spam / malware / phishing) in the SURBL, Google Safebrowsing or PhishTank blacklists.

The reason for including WOT reputation scores in our labeled dataset of malicious posts was two-fold. Firstly, as previously discussed, one of the goals of this work is to evaluate Facebook's current techniques to counter malicious content. Facebook partnered with WOT~\cite{Facebook-Developers:2011} to protect its users from malicious URLs (discussed further in \emph{Analysis and results} section). Evaluating the effectiveness of Facebook's use of WOT was one of the ways to achieve our goal. Secondly, when it comes to real world events, malicious entities tend to engage in spreading fake, untrustworthy and obscene content to degrade user experience~\cite{gupta2013faking,gupta20131}. This kind of information, despite being malicious, is not captured by blacklists like Google Safebrowsing and SURBL, since they do not fall under the more obvious kinds of threats like malware and phishing. WOT scores helped us to identify and tag such content.

%
%


\section{Analysis and results}

We now present our findings about the effectiveness of Facebook's current techniques of malicious content detection and the differences between malicious and legitimate content on Facebook. We then use these difference to identify a set of 42 features and apply standard machine learning techniques to automatically differentiate malicious content from legitimate content.

\subsection{Efficiency of Facebook's current techniques}

Facebook's immune system uses multiple URL blacklists to detect malicious URLs in real time and prevent them from entering the social graph~\cite{stein2011facebook}. Understandably, the inefficiency of blacklists to detect URLs at zero-hour makes this technique considerably ineffective~\cite{sheng2009empirical}. We tried to check if Facebook was taking any measures to overcome this drawback. To this end, we re-queried the Graph API for all malicious posts in our dataset and observed if Facebook was able to detect and remove malicious posts at a later point in time. We also studied the effectiveness of Facebook's partnership with WOT to protect its users from malicious URLs~\cite{Facebook-Developers:2011}.

\begin{figure}[!h]
\begin{center}
\includegraphics[scale=0.45]{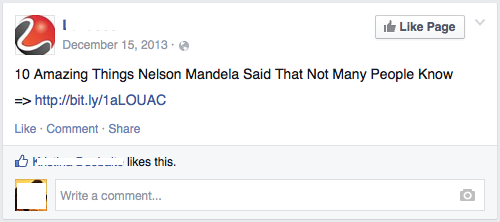}
\end{center}
\caption{One of the 7,296 malicious posts from our dataset which were not deleted by Facebook. We revisited this post after 11 months of being posted.}
\label{fig:spam_post}
\end{figure}

Upon re-querying the Graph API in November 2014, we found that only 3,921 out of the 11,217 (34.95\%) malicious posts had been deleted. It was surprising to note that almost two thirds of all malicious posts (65.05\%) which got past Facebook's real time detection filters remained undetected even after at least 4 months (between July 2014, when we stopped data collection, and November 2014, when we re-queried the API) from the date of post. Collectively, these posts had gathered \emph{likes} from 52,169 unique users and \emph{comments} from 8,784 unique users at the time we recollected their data in November 2014. Figure~\ref{fig:spam_post} shows one such malicious post from our dataset which went undetected by Facebook. The short URL in the post points to a scam website which misleads users into earning money by \emph{liking} posts on Facebook. Using the URL endpoint of the Graph API~\footnote{\url{https://developers.facebook.com/docs/graph-api/reference/v2.2/url}}, we also found that the 4,622 unique URLs present in the 11,217 malicious posts had been shared on Facebook over 37 million times. A possible reason for the continued existence of malicious posts could be that Facebook does not rescan a post once it passes through Facebook's content filter and into the social network.

\begin{figure}[!ht]
\begin{center}
\includegraphics[scale=0.45]{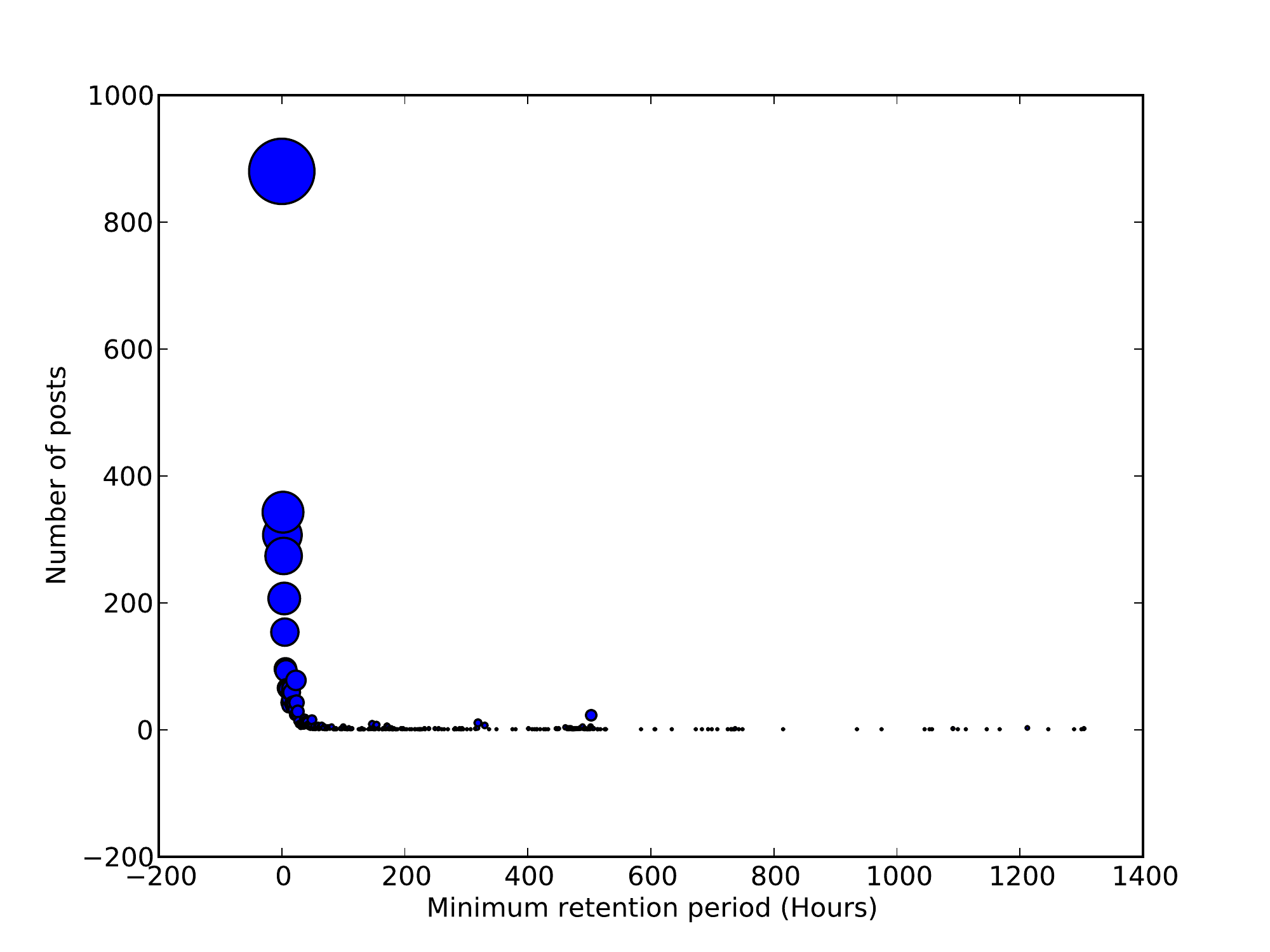}
\end{center}
\caption{Minimum retention period (in hours) of malicious posts in our dataset which were removed from Facebook. Approximately 51\% of the 3,921 posts were captured within the first 5 hours of being posted.}
\label{fig:time_post_graph}
\vspace{-5pt}
\end{figure}

For the 3,921 posts that were deleted, an interesting aspect to study was the amount of time it took for these posts to get removed. Although we were not able capture this information, we used the minimum retention period (time difference between the time of post creation and the time of post capture in our dataset) to estimate a lower bound on this value. We observed that the median time difference between a post's creation time and capture time was 4.64 hours ($\mu$ = 41.99 hours, $\sigma$ = 128.7 hours, min. = 1 second, max = 54 days). Figure~\ref{fig:time_post_graph} represents the minimum retention period of the 3,921 malicious posts in our dataset, which were not deleted by Facebook until November, 2014. 


To get an accurate approximation of the time it took for a malicious post to get deleted, the ideal approach would have been to re-query the Graph API for each post we collected, repeatedly after a fixed (small) interval of time. However, re-querying the API for all 4.4 million posts in our dataset repeatedly and periodically was a computationally expensive and infeasible task.

The above analysis suggests malicious content which goes undetected by Facebook's real time filters not only remains undetected for at least some time, but thrives on users' \emph{likes} and \emph{comments}. This increases the reach and visibility of malicious content and potentially exposes a much larger section of users than the attacker may have initially intended. A real time zero-hour detection technique can aid the identification and removal of such content, and can further improve Facebook's existing systems for malicious content detection. 

\paragraph{Partnership with Web of Trust}

\begin{figure}[!ht]
\begin{center}
\includegraphics[scale=0.35]{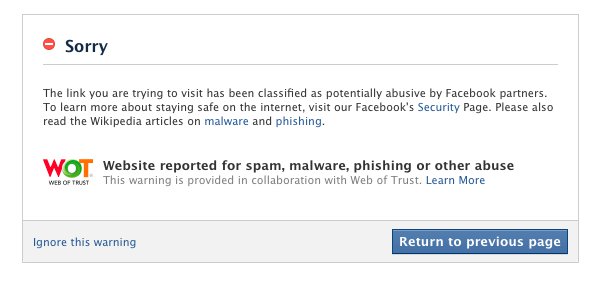}
\end{center}
\vspace{-5pt}
\caption{Warning page shown by Facebook whenever a user clicks on a link reported as abusive on Web of Trust. The user may chose to return or ignore the warning and visit the URL any way.}
\label{fig:fb_wot_warning}
\end{figure}

Facebook partnered with Web of Trust in 2011 to protect its users from malicious URLs~\cite{Facebook-Developers:2011}. According to this partnership, Facebook shows a warning page to the user whenever she clicks on a link which has been reported for spam, malware, phishing or any other kind of abuse on WOT (Figure~\ref{fig:fb_wot_warning}). To verify the existence of this warning page, we manually visited a random sample of 100 posts containing a URL marked as malicious by WOT, and clicked on the URL. Surprisingly, the warning page did not appear even once. We also noticed that over 88\% of all malicious URLs in our dataset (4,077 out of 4,622) were marked as malicious by Web of Trust. This highlights considerable inefficacy in Facebook's partnership with Web of Trust to control the spread of malicious URLs.

From the above analysis, it is evident that Facebook's existing techniques to combat the spread of malicious content through the social graph still have scope for improvement. In the next section, we highlight some key characteristics of the malicious content we found in our dataset. We present some important features which can be used to subdue the shortcomings of blacklist lookups and help Facebook in identifying malicious content from legitimate content efficiently and in real time.

\subsection{Key characteristics of malicious content}

We analyzed the malicious content in our dataset in three aspects; a) textual content and URLs, b) entities who post malicious content, and c) metadata associated with malicious content. We now look at all these three aspects individually in detail.

\subsubsection{Textual content and URLs}

From our dataset of 11,217 unique malicious posts, we first looked at the most commonly appearing posts. Similar to past work~\cite{gao2010detecting}, we found various \emph{campaigns} promoting a particular entity or event. However, campaigns in our dataset were very different than those discussed in the past. Table~\ref{tab:camps} shows the top 10 campaigns we found in our dataset of malicious posts. We found that most of the campaigns in our dataset were event specific, and targeted at celebrities and famous personalities who were part of the event. Although this seems fairly obvious because of our event based dataset, such campaigns reflect the attackers' preferences of using the context of an event to target OSN users. Attackers now prefer to exploit users' curiosity, instead of hijacking trends and posting unrelated content (like promoting free iPhone, illegal drugs, cheap pills, free ringtones, etc.) using topic specific keywords. 

\begin{table}
\small
\begin{tabular}{p{6.9cm}|p{0.6cm}}
\hline
Post Summary                                                                                        & Count \\ \hline
Sexy Football Worldcup - Bodypainting                                                 & 155      \\
10 Things Nelson Mandela Said That Everyone Should Know                 & 154      \\
Was Bishop Desmond Tutu Frozen Out of Nelson Mandela's Funeral?    & 105      \\
Nude images of Kate Middleton                                                             & 73        \\
The Gospel Profoundly Affected Nelson Mandela's Life After Prison     & 72         \\
Promotion of Obamacare (Affordable Care Act) through Nelson Mandela's death & 67 \\
Radical post about Nelson Mandela                                                      & 54         \\
Was Nelson Mandela a Christian?                                                          &  41        \\ 
R.I.P. Nelson Mandela: How he changed the world                                & 36         \\
Easy free cash                                                                                       & 29          \\

 \hline

\end{tabular}
\caption{Top 10 most common posts in our dataset of malicious posts. }
\label{tab:camps}
\vspace{-10pt}
\end{table}

We then looked at the various types of malicious posts present in our dataset. We found that the most common type of malicious posts were the ones with URLs pointing to adult content and incidental nudity, and marked unsafe for children (52.0\%) by Web of Trust. The second most common types of malicious posts comprised of negative (malware, phishing, scam, etc.) and questionable (misleading claims or unethical, spam, hate, discrimination, potentially unwanted programs, etc.) category URLs (45.2\%), closely followed by posts containing untrustworthy sources of information (38.22\%). Interestingly, only 325 malicious posts (2.9\%) advertised a phishing URL. This is a drastic drop as compared to the observations made by Gao et al. in 2010, where they found that over 70\% of all malicious posts in their dataset advertised phishing~\cite{gao2010detecting}. We also found that 18.4\% of the malicious posts in our dataset (2,064 posts out of 11,217) advertised one or more shortened URLs. Past literature has shown wide usage of shortened URLs to spread malicious content on microblogging platforms~\cite{Chhabra2011,antoniades2011we}. Short URLs have seen a significant increase in their usage mostly due to restriction of message length on OSNs like Twitter. However, given that this restriction on message length does not apply on Facebook, obfuscation of actual landing pages is, most likely, the primary reason behind usage of shortened URLs.

In addition to post categories, we also looked at the most common URL domains in our dataset. Table~\ref{tab:topdomains} shows the 10 most widely shared malicious and legitimate domains in our dataset. It is interesting to note that Facebook and YouTube constituted almost 60\% of all legitimate URLs shared during the 17 events. The remaining legitimate URLs largely belonged to news websites. On the contrary, malicious URLs were more evenly distributed across a mixture of categories including news, sports, entertainment, blogs, etc. Our dataset revealed that a large fraction of malicious content comprised of untrustworthy sources of information, which may have inappropriate implications in the real world, especially during events like elections, riots, etc. Most previous studies on detecting malicious content on online social networks have concentrated on identifying more obvious threats like malware and phishing~\cite{benevenuto2010detecting,grier2010spam,wang2010don}. There exists some work on studying trustworthiness of information on other social networks like Twitter~\cite{castillo2011information,gupta2012credibility}. However, to the best of our knowledge, no past work addresses the problem of identifying untrustworthy content on Facebook.

\begin{table*}[!ht]
\small
    \begin{tabular}{l|p{3.5cm}|p{3cm}|p{0.4cm}||l|p{2.1cm}|p{0.5cm}}
    \hline
    Malicious Domain & WOT categories & VirusTotal Category  & \% & Legitimate Domain & VirusTotal Category  & \% \\ \hline \hline
bizpacreview.com & Untrustworthy  & News & 5.60 & facebook.com & Social networks & 53.17  \\ \hline
9cric.com & Child unsafe  & Sports & 4.69  & youtube.com      & Social web     & 6.69  \\ \hline
imgur.com  & Child unsafe & Online Photos & 3.45 & cnn.com   & News  & 0.72     \\ \hline
allchristiannews.com  & Untrustworthy, Child unsafe, hate, discrimination  & News, Traditional religions    & 2.53       & bbc.co.uk          & News       & 0.58       \\ \hline
 worldstarhiphop.com & Child unsafe, adult content, gruesome or shocking  & Entertainment, Streaming media   & 2.42       & twitter.com            & Social networks    & 0.50       \\ \hline
 mobilelikez.com   & Untrustworthy, Child unsafe, spam  & Society and lifestyle & 2.21 & theguardian.com   & News   & 0.49       \\ \hline
liveleak.com  & Child unsafe, adult content, gruesome or shocking   & \raggedright News, Violence, Streaming media    & 2.01  & za.news.yahoo.com  & Search engine & 0.47   \\ \hline
25.media.tumblr.com   & Child unsafe, adult content  & Blogs and personal sites, adult content & 1.44  & dailymail.co.uk  & News, entertainment     & 0.39       \\ \hline
sensuelweb.com  & Adult content  & Adult content   & 1.27   & apps.facebook.com   & Social networks    & 0.37       \\ \hline
exopolitics.blogs.com  & Child unsafe, adult content   & Blogs / social networks & 1.21       & huffingtonpost.com & News & 0.31       \\ \hline
    \end{tabular}
\vspace{-10pt}
\caption{Top 10 malicious and legitimate domains and their VirusTotal categories (and Web of Trust categories for malicious domains) in our dataset. Most legitimate URLs shared on Facebook belonged to their own domain. Malicious domains were much more evenly spread.}
\label{tab:topdomains}
\vspace{-10pt}
\end{table*}


\subsubsection{Entities posting malicious content}

Having found a reasonable number of malicious posts in our dataset, we further investigated the entities (users / pages) who generated these malicious posts. We found that over 25\% of the entire 3.3 million unique entities in our dataset (approx. 0.85 million) posted at least one URL, and 0.24\% (7,962) unique entities posted at least one malicious URL (see Table~\ref{tab:descstats}). Content on Facebook is generated by two types of entities; \emph{users} and \emph{pages}. Pages are public profiles specifically created for businesses, brands, celebrities, causes, and other organizations, often for publicity. Unlike users, pages do not gain ``friends," but ``fans," people who choose to ``like" a page. In our dataset, we identified pages by the presence of \emph{category} field in the response returned by Graph API search~\cite{Facebook-Developers:2013} during the initial data collection process. The \emph{category} field is specific to pages and we manually verified that it was returned for all pages in our dataset. All remaining entities were users. We queried the Graph API~\footnote{\url{https://graph.facebook.com/<entity_id>}} for profiles of all the 3.3 million entities in our dataset in October 2014. During this process, \emph{gender} information for all users was acquired and changes in page \emph{category} (if any) were captured. We found that 10.2\% of the pages (39,843) had been deleted, and 8.7\% of the pages (33,794) had changed their category.

\begin{figure}[!ht]
     \begin{center}
        	\subfigure[]{%
            \label{fig:spammers}
            \includegraphics[scale=0.34]{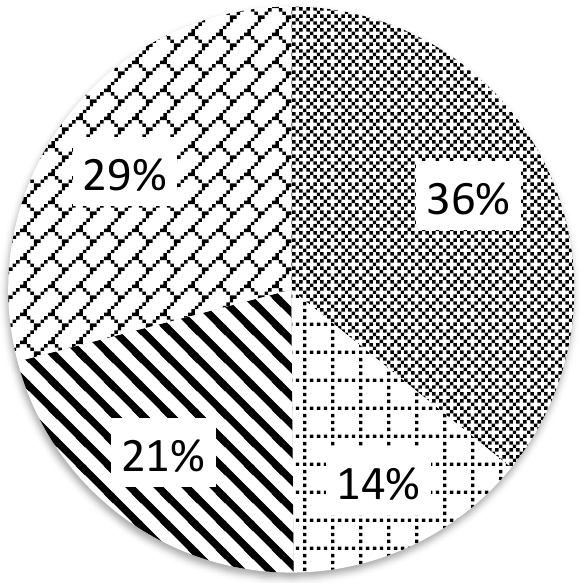}
        } 
        \subfigure[]{%
           \label{fig:non_spammers_urls}
           \includegraphics[scale=0.44]{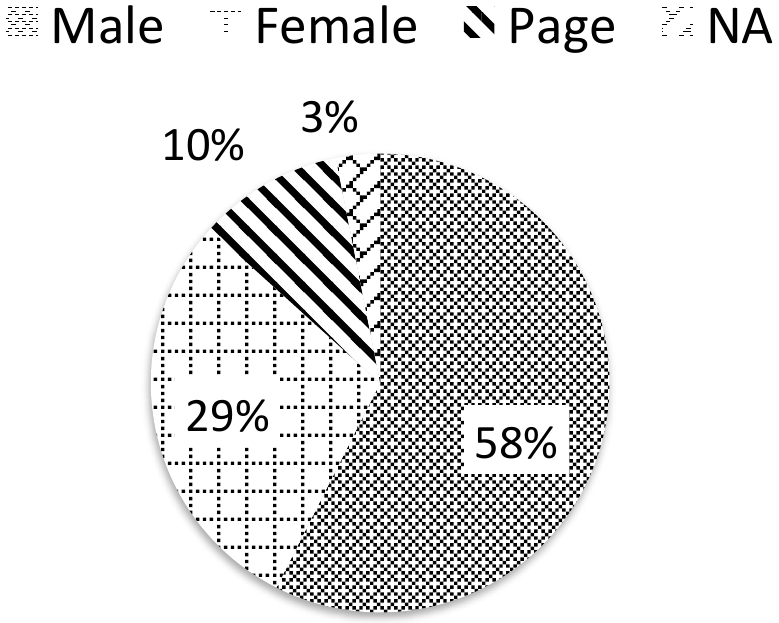}
        } 
\subfigure[]{%
           \label{fig:non_spammers_all}
           \includegraphics[scale=0.35]{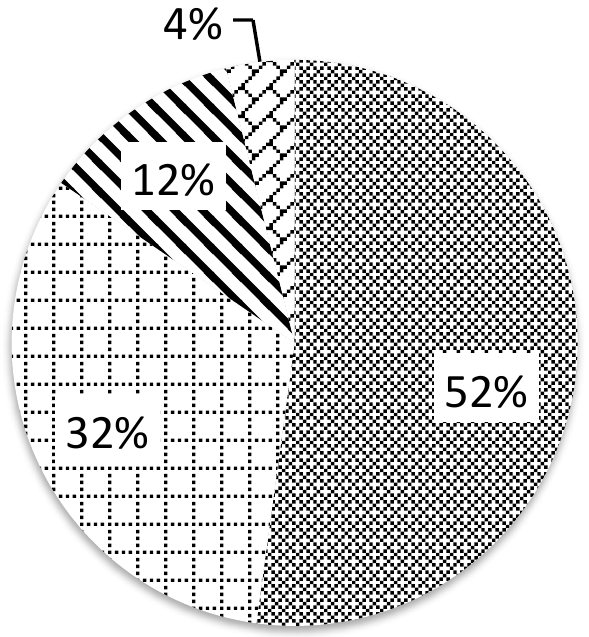}
        }
    \end{center}
\vspace{-5pt}
    \caption{Gender and category distribution of (a) Malicious entities (N=7,962); (b) Non malicious entities posting URLs (N=849,190); and (c) All non malicious entities (N=3,365,991) in our dataset.
     }%
   \label{fig:pie_gender_page}
\vspace{-10pt}
\end{figure}

Upon analyzing users, we found that for users who posted malicious URLs, the gender distribution was skewed towards the male population as compared to the gender distribution for legitimate users in our dataset (see Figure~\ref{fig:pie_gender_page}). For malicious users, the {\fontfamily{pcr}\selectfont male:female} ratio was {\fontfamily{qcr}\selectfont 1:2.41} (Figure~\ref{fig:spammers}), whereas for legitimate users posting one or more URLs, this ratio was {\fontfamily{qcr}\selectfont 1:2} (Figure~\ref{fig:non_spammers_urls}). The {\fontfamily{qcr}\selectfont male:female} distribution further dropped to {\fontfamily{qcr}\selectfont 1:1.62} for all legitimate users (Figure~\ref{fig:non_spammers_all}). We also found that pages were more active in posting malicious URLs as compared to non malicious URLs. Pages were observed to constitute 21\% (1,676 out of 7,962) of all malicious entities (Figure~\ref{fig:spammers}), while only 10\% of all legitimate URL posting entities were pages (Figure~\ref{fig:non_spammers_urls}). A similar percentage of pages (12\%) was found to constitute all legitimate entities in our dataset (Figure~\ref{fig:non_spammers_all}). We also found 43 verified pages and 1 verified user among entities who posted malicious content. The most common type of verified pages were radio station pages (12), website pages (5) and public figure pages (4). Combined together, the 43 verified pages had over 71 million \emph{likes}.

\begin{figure}[!ht]
\begin{center}
\includegraphics[scale=0.33]{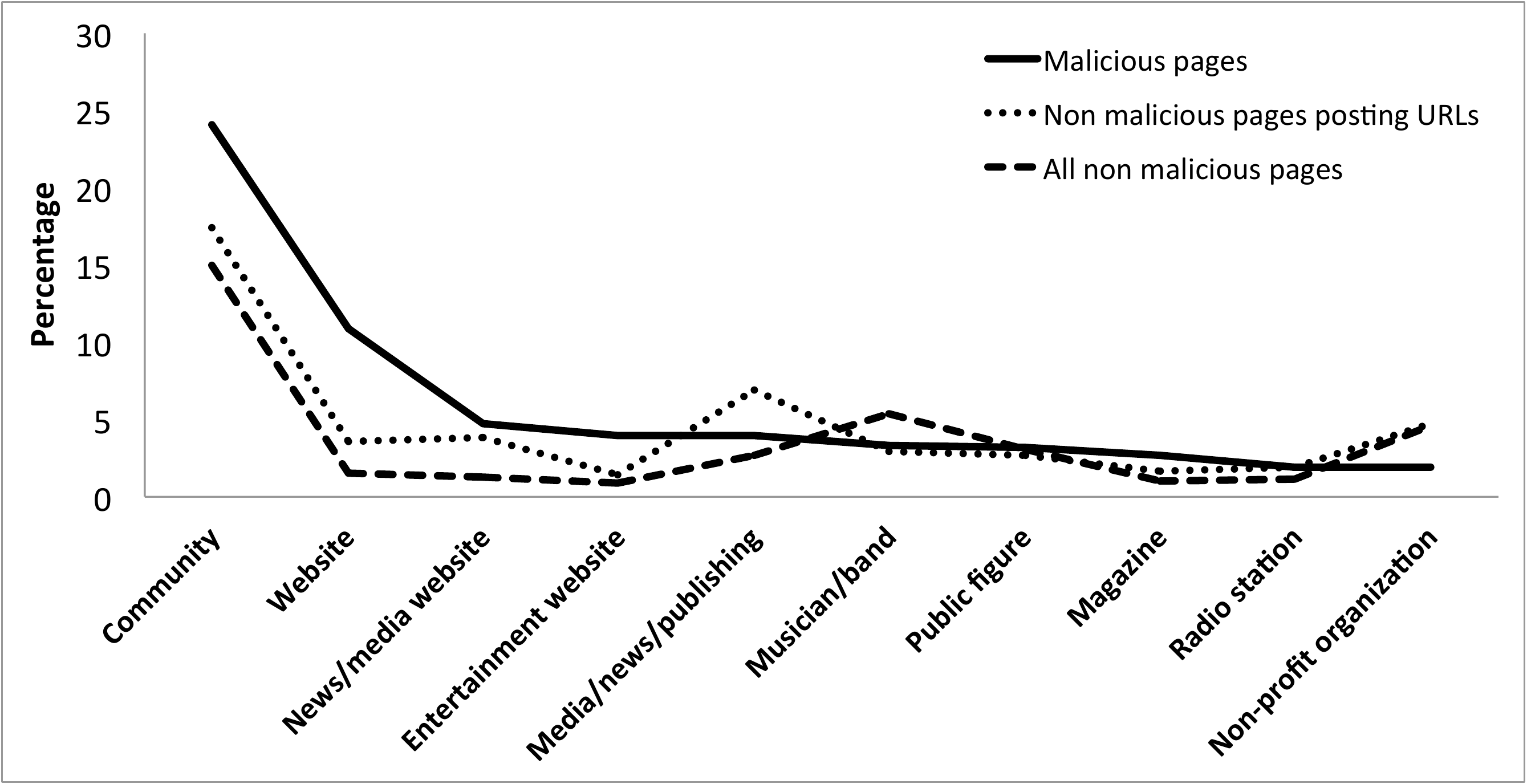}
\end{center}
\vspace{-10pt}
\caption{Top ten page categories among Facebook pages posting malicious content. We found similar category distribution among malicious and non-malicious pages.}
\label{fig:categories}
\vspace{-7pt}
\end{figure}

It is important to note that most of the past attempts at studying malicious content on Facebook did not capture content posted by pages, and concentrated only on users~\cite{ahmed2012mcl,gao2010detecting,stringhini2010detecting}. Malicious content originating from pages in our dataset brings out a new element of Facebook, which is yet to be addressed. Facebook limits the number of \emph{friends} a user can have, but there is not limit on the number of people who can subscribe to (\emph{Like}, in terms of Facebook terminology) a page. Content posted by a page can thus, have much larger audience than that of a user, making malicious content posted by pages potentially more widespread and dangerous than that posted by individual users. We found that in our dataset, pages posting malicious content had 123,255 \emph{Likes} on average (min. = 0, max. = 50,034,993), whereas for legitimate pages, the average number of \emph{Likes} per page was only 45,812 (min. = 0, max. = 502,938,006). 
Upon further investigation, we found high similarity between the \emph{categories} of the most famous pages posting legitimate and malicious content. We found that the 10 most famous categories of pages posting malicious content were also among the most famous categories of pages posting legitimate content (Figure~\ref{fig:categories}). 

\subsubsection{Metadata}

\begin{figure}[!ht]
\begin{center}
\includegraphics[scale=0.4]{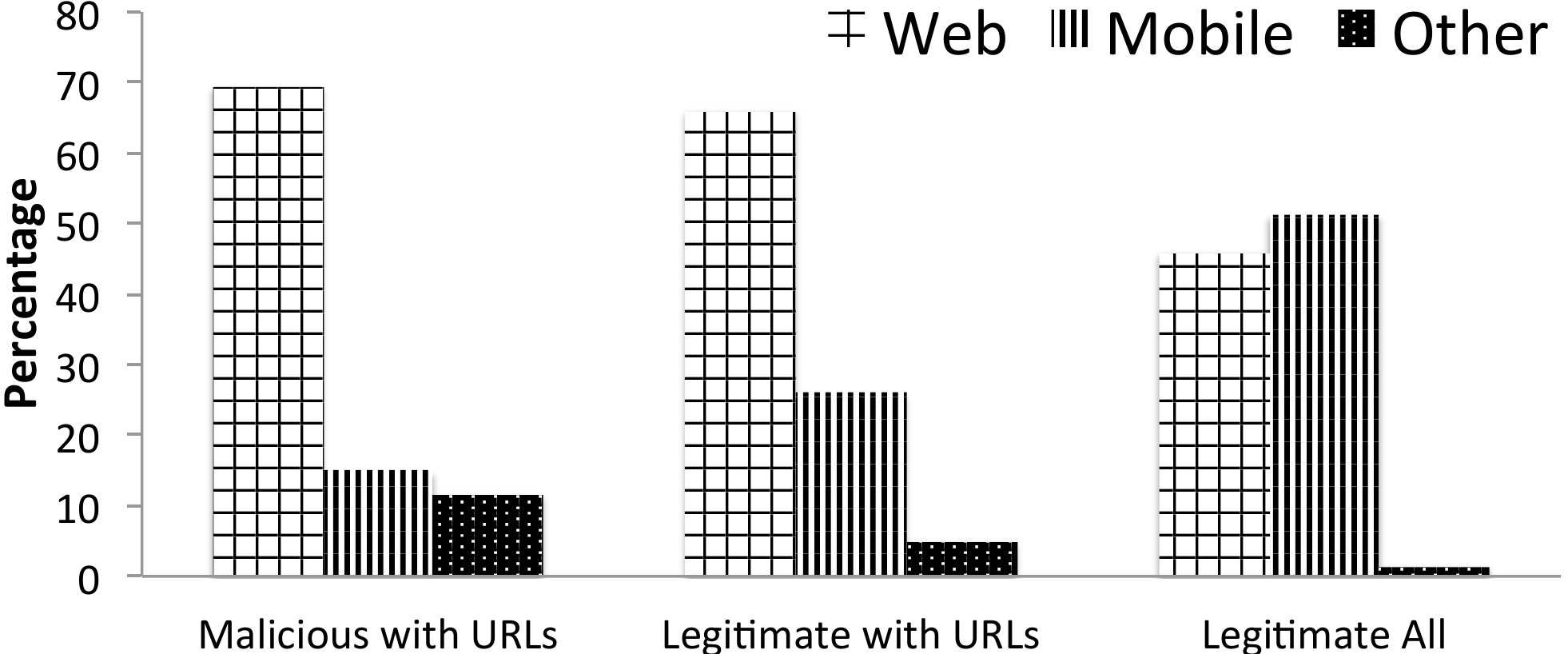}
\end{center}
\vspace{-10pt}
\caption{Sources of malicious content, legitimate content with URLs, and all legitimate content. Mobile platforms were preferred over web for posting legitimate content.}
\label{fig:apps}
\vspace{-5pt}

\end{figure}

There are various types of metadata associated with a post, for example, application used to post, time of post, type of post (picture / video / link), location etc. Metadata is a rich source of information that can be used to differentiate between malicious and legitimate users. Figure~\ref{fig:apps} shows the distribution of the top 25 applications (web / mobile / other) used to post content in our dataset.~\footnote{The top 25 applications were used to generate over 95\% of content in all three categories we analysed.} We observed that over 51\% of all legitimate content was posted through mobile apps. This percentage dropped to below 15\% for malicious content. Third party and custom applications (captured in ``Other" in Figure~\ref{fig:apps}) were used to generate 11.5\% of all malicious content in our dataset as compared to only 1.4\% of all legitimate content being generated by such applications. This behavior reflects that malicious entities make use of web and third party applications (possibly for automation) to spread malicious content, and can be an indicator of malicious activity. Legitimate entities, on the other hand, largely resort to standard mobile platforms to post.

Although Facebook has more web users than mobile users~\cite{Facebook:2014}, our observations may be biased towards mobile users due to our event specific dataset. As described in the Data section, our data was collected during 17 real world events. Past literature has shown high social network activity through mobile devices during such events~\cite{gupta20131}.

We also observed significant difference in the content types that constituted malicious and legitimate content. Over 50\% of legitimate posts containing a URL were photos or videos whereas this percentage dropped to below 6\% for malicious content. A large proportion of these photos and videos were uploaded on Facebook itself. This was one of the main reasons for facebook.com being the most common legitimate domain in our dataset (see Table~\ref{tab:topdomains}). We used these, and some other features to train multiple machine learning algorithms for automatic detection of malicious content. The results of our experiments are presented in the next section.


%
%
%

\section{Detecting malicious content automatically} \label{sec:ml}

Past efforts for automatic detection of spam and malicious content on Facebook largely focus on detecting campaigns~\cite{gao2012towards,gao2010detecting}, and rely heavily on \emph{message similarity} features to detect malicious content~\cite{rahman2012efficient}. Researchers using this approach have reported consistent accuracies of over 80\% using small feature sets comprising of under 10 features. However, this approach is ineffective in zero-hour detection since the aforementioned models require to have seen similar spam messages in the past. To overcome this inability, we propose an extensive set of 42 features (see Table~\ref{tab:features}) to detect malicious content, excluding features like message similarity, likes, comments, shares etc., which are absent at zero-hour. We group these 42 features into four categories based on the their source; Entity (E), Text content (T), Metadata (M) and Link (L).

\begin{table}[!h]
\small
    \begin{tabular}{p{1.2cm}|p{6.4cm}}
    \hline
    Source   & Features                                                                                                                                                                                                                                                                                                                             \\ \hline
    Entity (9)  & is a page / user, gender, page category, has username, username length, name length, num. words in name, locale, likes on page                                                                                                                                                                                                       \\ \hline
    Text content (18)  & Presence of !, ?, !!, ??, emoticons (smile, frown), num. words, avg. word length, num. sentences, avg. sentence length, num. English dictionary words, num. hashtags, hashtags per word, num. characters, num. URLs, num. URLs per word, num. uppercase characters, num. words / num. unique words \\ \hline
    Metadata (8) & App, has FB.com URL, has \emph{message}, has \emph{story}, has \emph{link}, has \emph{picture}, type, \emph{link} length                                                                                                                                   \\ \hline
    Link (7)     & has HTTP / HTTPS, hyphen count, paramters count, parameter length, num. subdomains, path length                                                                                                                                                                                                                                      \\ \hline

    \end{tabular}
\vspace{-10pt}
\caption{Features used for machine learning experiments. We extracted features from four sources, viz. entity, content, metadata, and link.}
\label{tab:features}
\vspace{-5pt}
\end{table}

\begin{figure}[!ht]
     \begin{center}
        	\subfigure[]{%
            \label{fig:rf_acc_feats}
            \includegraphics[scale=0.28]{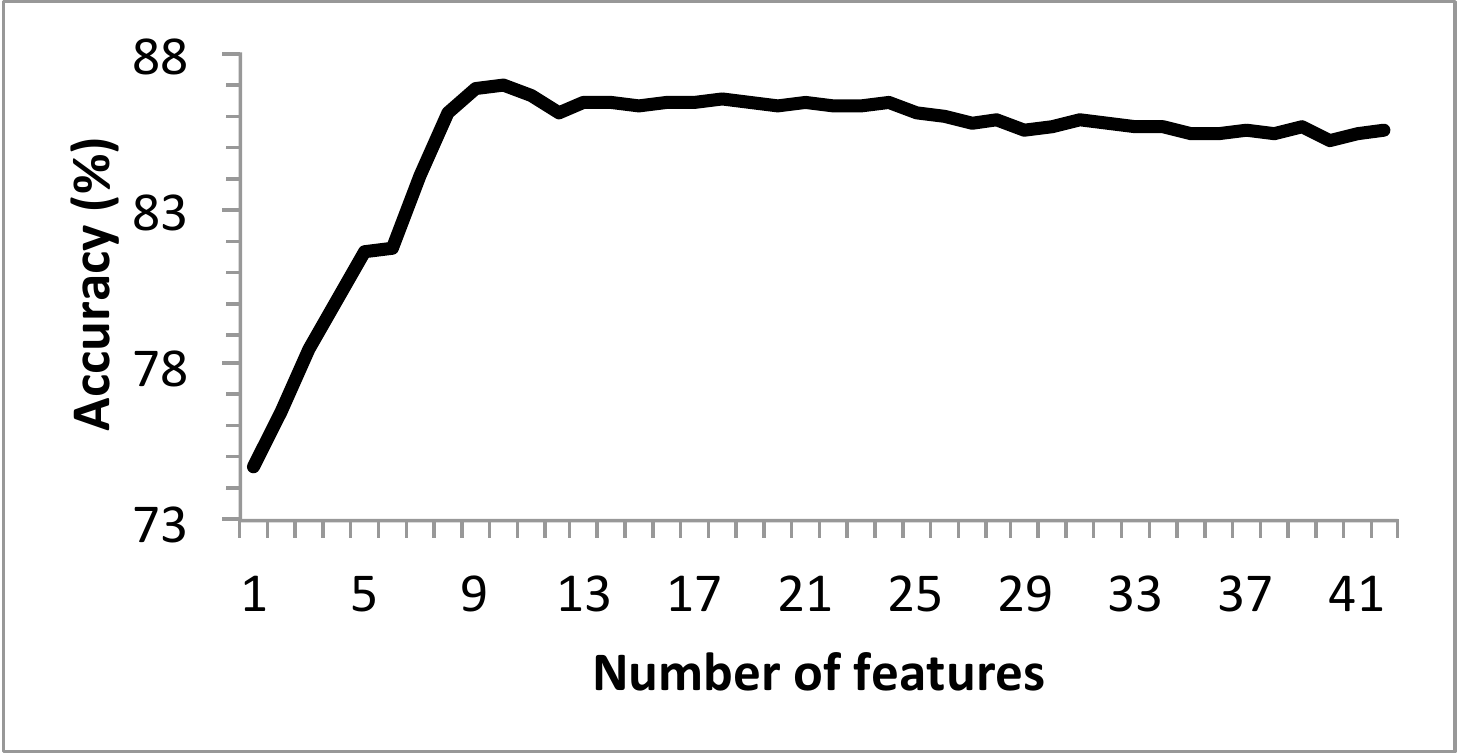}
        }
\subfigure[]{%
           \label{fig:non_spammers_all}
	  \includegraphics[scale=0.21]{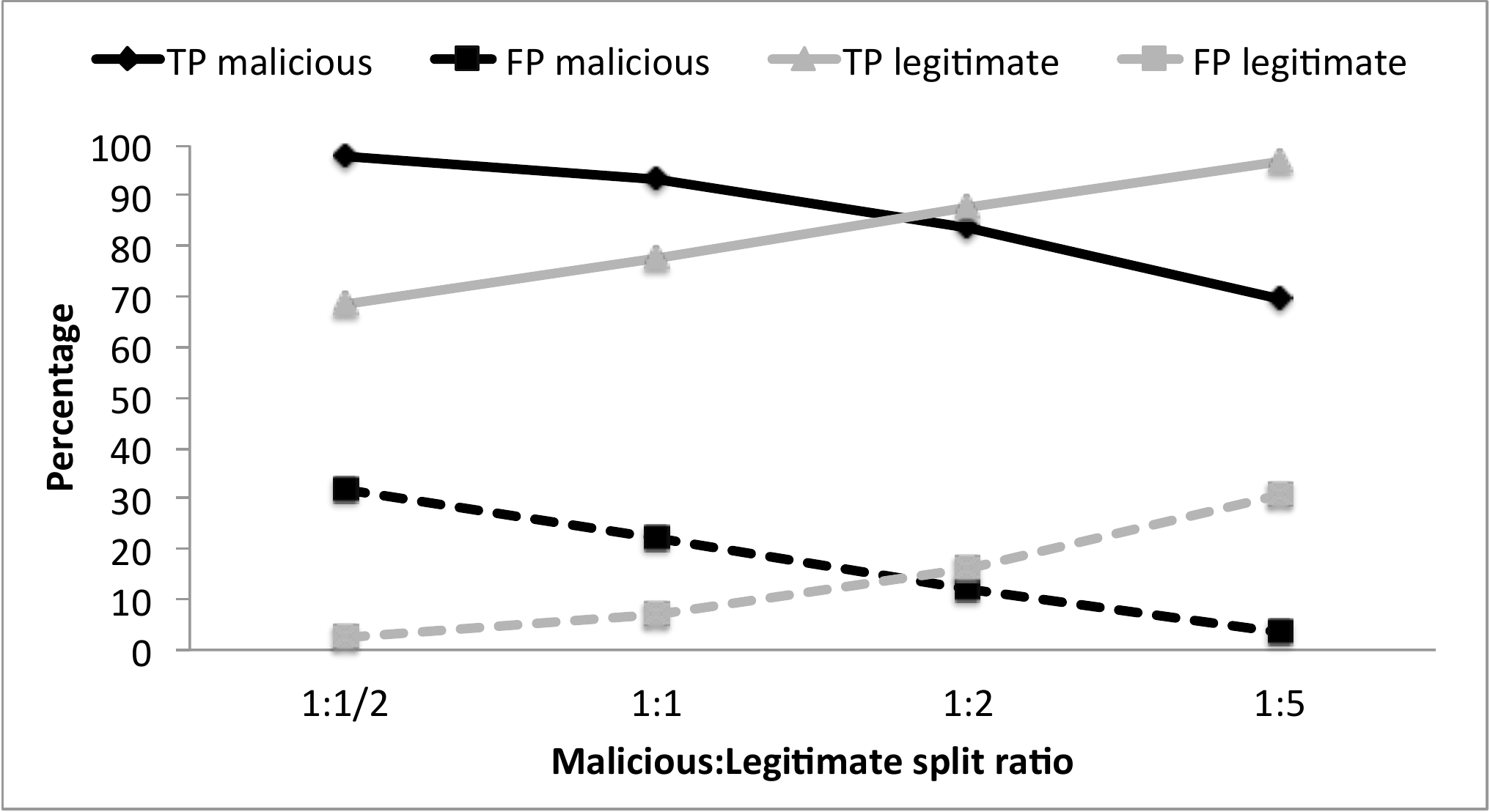}
        }
    \end{center}
\vspace{-10pt}
    \caption{(a) Accuracy values of the Random Forest classifier for 1 through 42 features. Accuracy peaked to 86.9\% at top 10 features. (b) True positive and false positive rates of malicious and legitimate classes with different sizes of the training set.
     }%
   \label{fig:tpfp}
\vspace{-5pt}
\end{figure}

We trained four classifiers using 11,217 unique malicious posts as the positive class and 11,217 unique legitimate posts, randomly drawn from the 1,210,920 unique legitimate posts containing one or more URLs (see Table~\ref{tab:descstats}) as the negative class. All experiments were performed using Weka~\cite{Hall2009}. A 10-fold cross validation on this training set yielded a maximum accuracy of 86.9\% using the Random Forest classifier. Table~\ref{tab:classifyresults} describes the results in detail. We also performed the classification experiments using the four category features (E, T, M, and L) separately, and observed that link (L) features performed the best, yielding an accuracy of 82.3\%. A combination of all four category features, however, outperformed the individual category scores, signifying that none of the category features individually could identify malicious posts as accurately as their combination. We also recorded the accuracy results for the Random Forest classifier by varying the number of features according to their information gain values (see Figure~\ref{fig:rf_acc_feats}). We calculated the accuracy for 1 through all 42 features, adding features one by one in decreasing order of their information gain value, and found that the accuracy peaked at the top 10 features. All four classifiers achieved higher accuracy when trained on the top 10 features, as compared to accuracy when trained on all 42 features (see Table~\ref{tab:classifyresults}). Table~\ref{tab:top10} shows the information gain value and source of the top 10 features.

\begin{table}[!h]
\begin{center}
\small
    \begin{tabular}{l|p{0.5cm}|p{0.5cm}|p{0.5cm}|p{0.5cm}|p{0.5cm}|p{0.7cm}}
    \hline
    Feature Set         &    E     &     T    &    M     &     L     &  All    &    Top10 \\ \hline
    Naive Bayesian   &  58.9  &  52.0  &  75.0   &  66.3  & 58.8  & 74.3      \\
    Decision Tree     &  {\bf 63.8}  &  65.4  &  80.8   &  82.0  & 85.0  & 85.8      \\
    Random Forest   &  63.6  &  {\bf 65.6}  & {\bf 80.9}   &  {\bf 82.3}  & {\bf 85.5} & {\bf 86.9}   \\
    AdaBoost            &  59.5  &  62.8  &  76.5   &  71.8  & 76.8  & 77.4      \\ \hline
    \end{tabular}
\end{center}
\vspace{-10pt}
\caption{Ten-fold cross validation accuracies for four classifiers over six different feature sets. 
}
\label{tab:classifyresults}
\vspace{-10pt}
\end{table}


\begin{table}[!ht]
\small
\begin{center}
    \begin{tabular}{l|l|l}
    \hline
    Feature                           & Source   & Info. Gain \\ \hline
    Presence of Facebook.com URL      & Metadata & 0.240            \\
    Post type                         & Metadata & 0.219            \\
    Length of parameter(s) in URL     & Link     & 0.216            \\
    Application used to post          & Metadata & 0.209            \\
    Length of \emph{link} field                    & Metadata & 0.201            \\
    Number of parameters in URL       & Link     & 0.178            \\
    Number of sub-domains in URL      & Link     & 0.110            \\
    Length of URL path (after domain) & Link     & 0.093            \\
    Number of hyphens in URL          & Link     & 0.084            \\
    Presence of \emph{story} field           & Metadata & 0.071            \\ \hline
    \end{tabular}
\vspace{-5pt}
\caption{Source and information gain value of the top 10 features.}
\label{tab:top10}
\end{center}
\vspace{-10pt}
\end{table}

We performed further experiments to observe the change in true positive rate, false positive rate and accuracy values with change in training dataset sizes. The Random Forest classifier was used for these experiments since it gave the highest accuracy amongst the four classifiers we used in the previous experiment. We used all 42 features to train the classifier in this experiment. Keeping the size of the positive class constant (11,217 instances), the size of the training set was varied by varying the size of the negative class instances in the ratio 1:1/2, 1:1, 1:2 and 1:5. This yielded 5,609, 11,217, 22,434 and 56,085 negative class instances consecutively, randomly drawn from 1,210,920 unique legitimate posts containing one or more URLs (see Table~\ref{tab:descstats}). Figure~\ref{fig:tpfp} shows the results of this experiment. We were able to achieve a maximum true positive rate of 97.7\% for malicious posts using the 1:1/2 split. The false positive rate dropped to a lowest of 3.4\% for 1:5 split. As we increased the size of the negative class, both the true positive and false positive rates for the malicious class constantly decreased. The 1:1 split yielded the lowest average false positive rate across both classes.

To check the effectiveness of our model over time, we collected test data about the Ebola outbreak in Africa during August - October 2014. We collected a total of 59,179 posts containing URLs, out of which, 3,248 post were found to be malicious.~\footnote{We used the same methodology to find malicious posts as we did for the 17 events in our training data.} Our final test set consisted of 6,496 posts (3,248 malicious and 3,248 randomly picked legitimate posts). Evaluating our balanced model (trained on 11,217 malicious and 11,217 legitimate posts) on this test set revealed a significant drop in true positive rate, from 93.2\% to 81.6\%. This value is, however, a slight improvement over previously reported true positive rates for spam detection techniques over time~\cite{gao2012towards}.

\subsection{Comparison with past work}

One of the few studies on detecting malicious content on Facebook using a dataset bigger than ours, was conducted by Gao et al., where researchers used a clustering approach to identify spam campaigns~\cite{gao2010detecting}. Although authors reported a high true positive rate of 93.9\%, there was no estimation of the amount of malicious posts that this approach missed (false negatives). To this end, we applied the same clustering technique and threshold values used by Gao et al. on our dataset to get an estimate of the false negatives of their clustering approach. Since our entire dataset was already labeled (as opposed to Gao et al.'s dataset), we did not apply clustering on our entire dataset to find malicious posts. Instead, we applied clustering only on malicious posts in our dataset, and compared how many of those clusters met the \emph{distributed} and \emph{bursty} threshold values previously used ($>$5 users per cluster, and $<$90 minutes median time between consecutive posts respectively). Applying clustering on our 11,217 malicious posts yielded a total of 4,306 clusters. Out of these, only 183 clusters (containing 4,294 posts) met the \emph{distributed} and \emph{bursty} thresholds, yielding a high false negative rate of 61.7\%. These results indicate that our machine learning models perform considerably better, and are able to detect more than double the amount of malicious posts as compared to existing clustering techniques. We were unable to compare our results with other previous work due to two major reasons; a) absence of features like \emph{likes, comments, message similarity} etc. at zero-hour (used by Rahman et al.), and b) public unavailability of features like number of friends, message sent, friend choice, active friends, page likes etc. (used by Stringhini et al.~\cite{stringhini2010detecting} and Ahmed et al.~\cite{ahmed2012mcl}). 

\section{REST API and browser plug-in}

To provide a real world solution for the problem of detecting malicious content on Facebook, we built a REST based API (Application Programming Interface) using the Random Forest classifier trained our labeled dataset. The API is publicly accessible at \url{http://multiosn.iiitd.edu.in/fbapi/endpoint/?fid=<Facebook_post_ID>} and can be queried by sending a HTTP GET request along with a Facebook post ID. Due to Facebook's API limitations, our API currently works only for public posts which are accessible through Facebook's Graph API. Our API fetches the post and entity profile information using Facebook's Graph API and generates a feature vector, which is subjected to a pre-trained classifier. The label returned by the classifier is output by the API in JSON format along with the original Facebook post ID.

\begin{figure}[!ht]
\begin{center}
\includegraphics[scale=0.25]{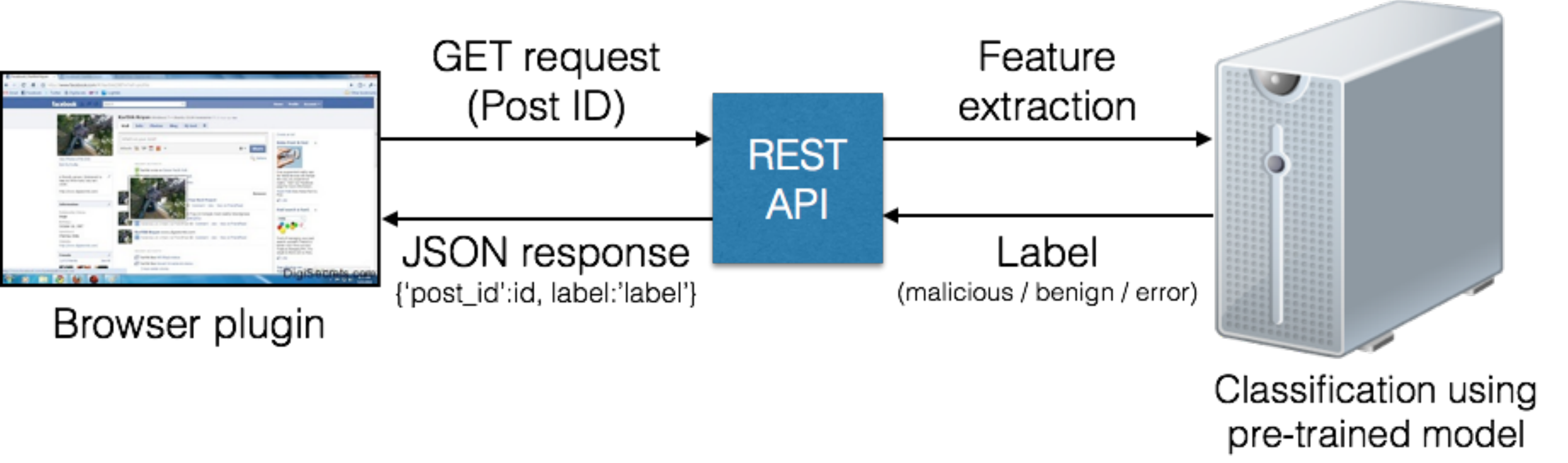}
\end{center}
\vspace{-10pt}
\caption{Flow diagram for our browser plug-in.}
\label{fig:plugin_arch}
\vspace{-10pt}
\end{figure}

\begin{figure}[!ht]
\begin{center}
\includegraphics[scale=0.36]{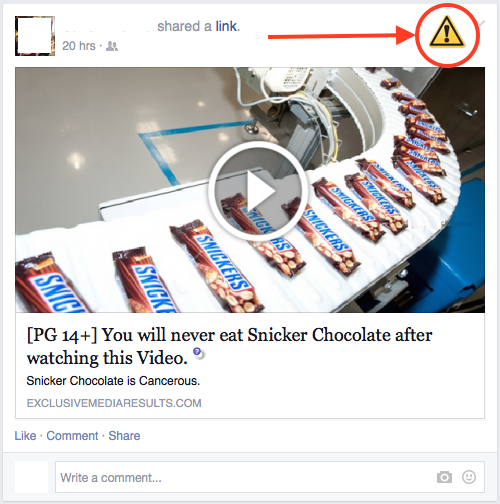}
\end{center}
\vspace{-10pt}
\caption{Screenshot of a malicious Facebook post labeled by our browser plug-in.}
\label{fig:plugin}
\vspace{-5pt}
\end{figure}

For better utility of our REST API, we built a plug-in for Google Chrome browser. Figure~\ref{fig:plugin_arch} shows the flow diagram representing the communication flow between the browser plug-in, REST API, and the classifier. Once installed and enabled, this plug-in loads whenever a user opens her Facebook page on her Google Chrome browser, and extracts the post IDs of all public posts in the user's newsfeed. The post IDs are then sent to the REST based API. If the API returns the label \texttt{malicious} for a post, the plug-in marks the post with an ``alert'' symbol (Figure~\ref{fig:plugin}).

\section{Limitations}

We could not find a way to claim that our dataset is representative of the entire Facebook population. Facebook does not provide any information about what percentage of public posts is returned by Graph API search. However, to the best of our knowledge, our dataset of 4.4 million public posts and 3.3 million users is the biggest dataset in literature, collected using Facebook APIs.

We understand that the WOT ratings that we used to create our labeled dataset of malicious posts are obtained through crowd sourcing, and may suffer biases. However, WOT states that in order to keep ratings more reliable, the system tracks each user's rating behavior before deciding how much it trusts the user. In addition, the meritocratic nature of WOT makes it far more difficult for spammers to abuse, because bots will have a hard time simulating human behavior over a long period of time. With these measures taken by WOT to control bias, we believe that it is safe to assume the validity of our labeled dataset of malicious posts and hence, our results.


\section{Conclusion}

OSNs witness large volumes of content during real world events, providing malicious entities a lucrative environment to spread scams, and other types of malicious content. We studied content generated during 17 such events on Facebook, and found substantial presence of malicious content which evaded Facebook's existing immune system and made it to the social graph. We observed characteristic differences between malicious and legitimate posts and used them to train machine learning models for automatic detection of malicious posts. Our extensive feature set was completely derived from public information available at zero-hour, and was able to detect more than double the number of malicious posts as compared to existing spam campaign detection techniques. Finally, we deployed a real world solution in the form of a REST based API and a browser plug-in to identify malicious Facebook posts in real time. In future, we would like to test the performance and usability of our browser plug-in. We would also like to investigate Facebook pages spreading malicious content in further detail. Further, we intend to study malicious Facebook posts which do not contain URLs.


%
\bibliographystyle{aaai}
\bibliography{/Users/prateekdewan/Dropbox/Super_BibTex_Collection/Prateek}  
\end{document}